\documentclass[twocolumn]{openjournal} 

\usepackage{lipsum}

\usepackage{xcolor}
\usepackage{textgreek}
\usepackage[utf8]{inputenc}
\usepackage[english]{babel}
\usepackage[breaklinks,colorlinks,citecolor=blue,urlcolor=blue]{hyperref}

\newcommand\blfootnote[1]{%
	\begingroup
	\renewcommand\thefootnote{}\footnote{#1}%
	\addtocounter{footnote}{-1}%
	\endgroup
}
	
\usepackage{color,colortbl}
\usepackage{tensind}
\tensordelimiter{?}
\DeclareGraphicsExtensions{.bmp,.png,.jpg,.pdf}
\usepackage{verbatim}
\usepackage[normalem]{ulem}
\usepackage{orcidlink}
\usepackage{soul}

\begin{document}
\title{Closure Invariants  for Polarised Radio Interferometric Observations: \\a graph theoretical approach}

\author{Vinay Kumar\orcidlink{0009-0008-2158-3774}$^{1\star}$}

\author{Rajaram Nityananda$^{1\dag}$}

\author{Joseph Samuel\orcidlink{0000-0003-2471-8714}$^{1, 2\ddag}$}

\affiliation{$^{1}$International Centre for Theoretical Sciences, Tata Institute of Fundamental Research, Bengaluru 560089, India}
\affiliation{$^{2}$Raman Research Institute, Bengaluru 560080, India}

\begin{abstract}
    Aperture synthesis observations with full polarisation have long been used to study  the magnetic fields  of synchrotron emitting  sources.
Recently proposed closure invariants give us a powerful method for extracting  information from measured visibilities which are corrupted by antenna and polarisation  dependent gains.
In this paper, a formalism developed earlier for complete graphs (where all visibilities are available)
is extended to incomplete graphs. The formalism provides a complete and independent set of closure
invariants from the measured visibilities in a general situation where not all visibilities are available.  We then  show in a  simulated, quasi-realistic  case
that the  invariants developed  here  contain  usable information even in the presence of  noise. 
\end{abstract}


	\blfootnote{\hspace{-0.5 cm} $\star$ \href{mailto:vinay.kumar@icts.res.in}{vinay.kumar@icts.res.in} \\
		$\dag$ \href{mailto:rajaram.nityananda@icts.res.in}{rajaram.nityananda@icts.res.in}
		\\
		$\ddag$ \href{mailto:sam@icts.res.in}{sam@icts.res.in}}

\maketitle

\section{Introduction}%
  
      Full polarisation observations with the Event Horizon Telescope have revealed the magnetic field structure of the innermost accretion flow around the M87 black hole (\cite{Akiyama_2021}). The principle underlying such observations
    is described in the standard text  (\cite{TMS}) and is summarised below.
     The measured electric fields  $\mathbf{E}_a$ at a given antenna `$a$'  are $2\times1$  complex matrices which are related to the true fields $\mathbf{T}_a$ by $2\times2$ complex `gain' matrices $\mathbf{G}_a$, i.e $\mathbf{E}_a=\mathbf{G}_a \mathbf{T}_a$. The measured correlations $\mathbf{M}_{ab}=\langle \mathbf{E}_a \mathbf{E}_b^\dagger\rangle $ are thus modified versions of the true correlations $\mathbf{S}_{ab}=\langle \mathbf{T}_a \mathbf{T}_b^\dagger\rangle$ given by the matrix product $\mathbf{M}_{ab}=\mathbf{G}_a \mathbf{S}_{ab} \mathbf{G}_b^\dagger$ . Here $\langle f\rangle$ is our notation for the  average of $f$ over the observation time.

      An important role is played in the imaging process by `closure quantities'.  These  are constructed from products of measured correlations around closed loops of antennas designed to cancel all the gains, rendering them invariant with respect to any changes in gains.  These have   long  been known for the  single polarisation case (\cite{TMS}), but were only recently introduced for full polarisation by \cite{Broderick_2020}, and in full generality  by \cite{Nithya} and  \cite{Samuel} for the case when all correlations are measured. We regard the antenna array as a graph, with antennas as vertices, and  measured baselines as edges. The existing  formalism is for the `complete graph' where every pair of vertices is joined by an edge. 
     In this work, we extend the framework by constructing a complete and independent set of closure invariants in the more general case of an incomplete graph, i.e when not all correlations between  pairs are available. In a heterogeneous array, this could happen for correlations between the smallest antennas. 
 We also explore the noise properties of the invariants, which are  nonlinear functions of the correlations. This step is necessary to construct likelihood functions for models to compare  to measurements.

The  principle of gain cancellation is that along  any circuit, we construct a matrix product around a closed loop.   A measured 
   correlation matrix    $\mathbf{M}_{ab}$ is followed with the inverse adjoint of $\mathbf{M}_{bc}$, i.e the  matrix ${(\mathbf{M}_{bc}^{\dagger })^{-1}}$ which we denote,
   for brevity, by  ${\widehat{\mathbf{M}}}_{bc}$. This ensures that the terminal $\mathbf{G}_b ^\dagger$ of $\mathbf{M}_{ab} $ is cancelled
   by the initial $\mathbf{G}_b^{\dagger^{-1}}$ of  ${\widehat{\mathbf{M}}}_{bc}$. Coming back to the starting point `a' after an even number of baselines, we now only have an initial $\mathbf{G}_a $ and a final $\mathbf{G}_a^{-1} $ We call such a product around an even number of baselines a `covariant'.  To construct invariants, we can take the trace and the determinant of this product, both of which are independent of $\mathbf{G}_a$ (\cite{Broderick_2020}).  The novel contribution of \cite{Samuel} which was used  to find the complete, independent set of invariants  was to introduce products of measured correlation matrices over loops with an {\it odd} number
   of baselines ({\it e.g} triangles), named as `advariants'. These start with a $\mathbf{G}_a$ and end with a $\mathbf{G}_a^\dagger$. Invariants were constructed from these advariants using properties of Lorentz transformations, as explained below.

   \begin{figure}
    \centering
    \begin{minipage}{.4732\columnwidth}
     \includegraphics[width = \columnwidth]{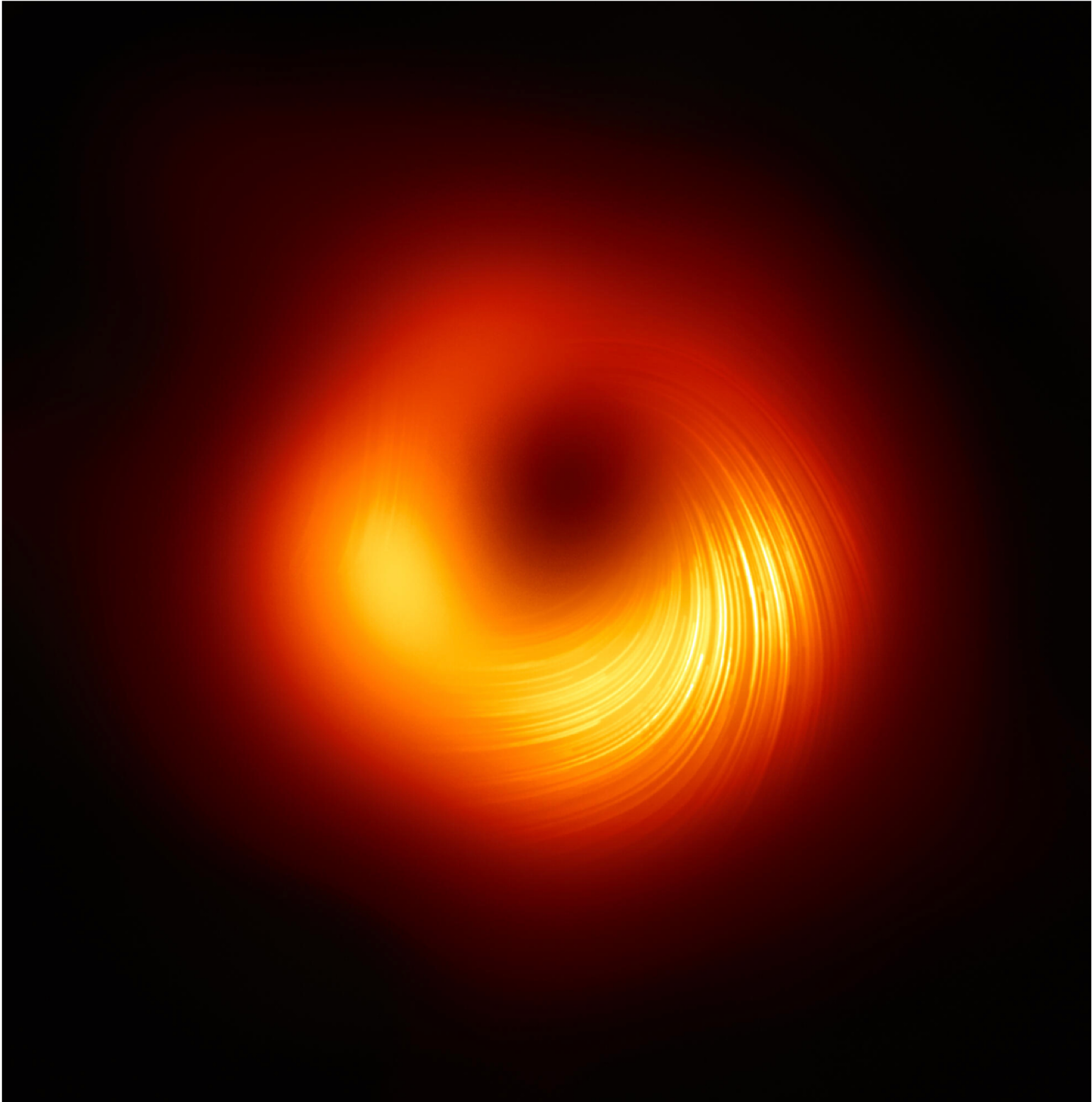}
    \end{minipage}%
    \begin{minipage}{.48\columnwidth}
      \includegraphics[width = \columnwidth]{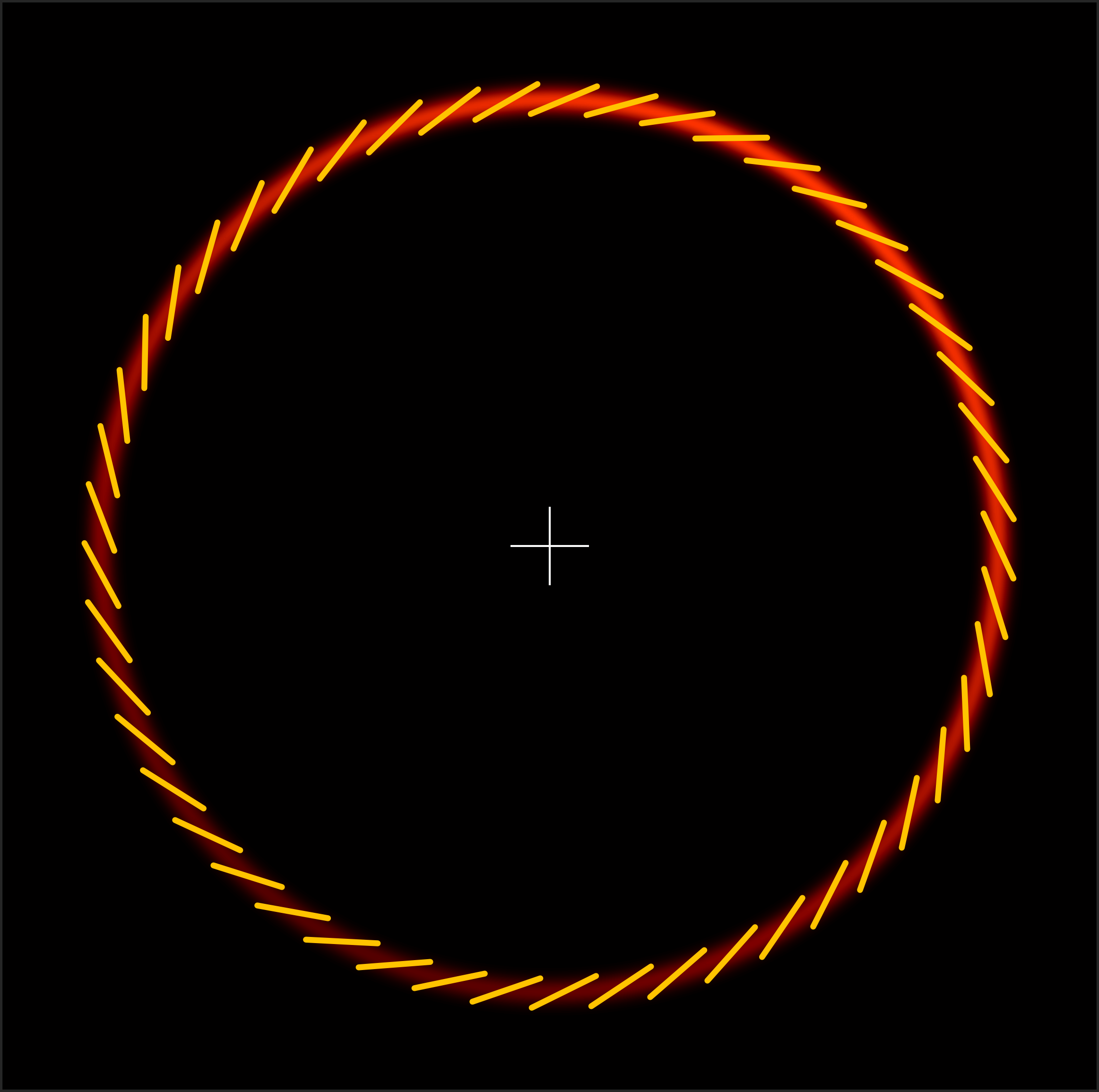}
     \end{minipage}
    \label{Rings}
    
    \caption{The image on the left shows the first polarised image of M87$^*$ accretion flow taken by the EHT collaboration, adapted from under a CC-BY 4.0 license. On the right is the schematic diagram of the ring image used in our numerical work to emulate features of the figure on the left.}
    \end{figure}

   \section{The incomplete graph case}%
 
\begin{figure}
\centering

 \includegraphics[width = \columnwidth]{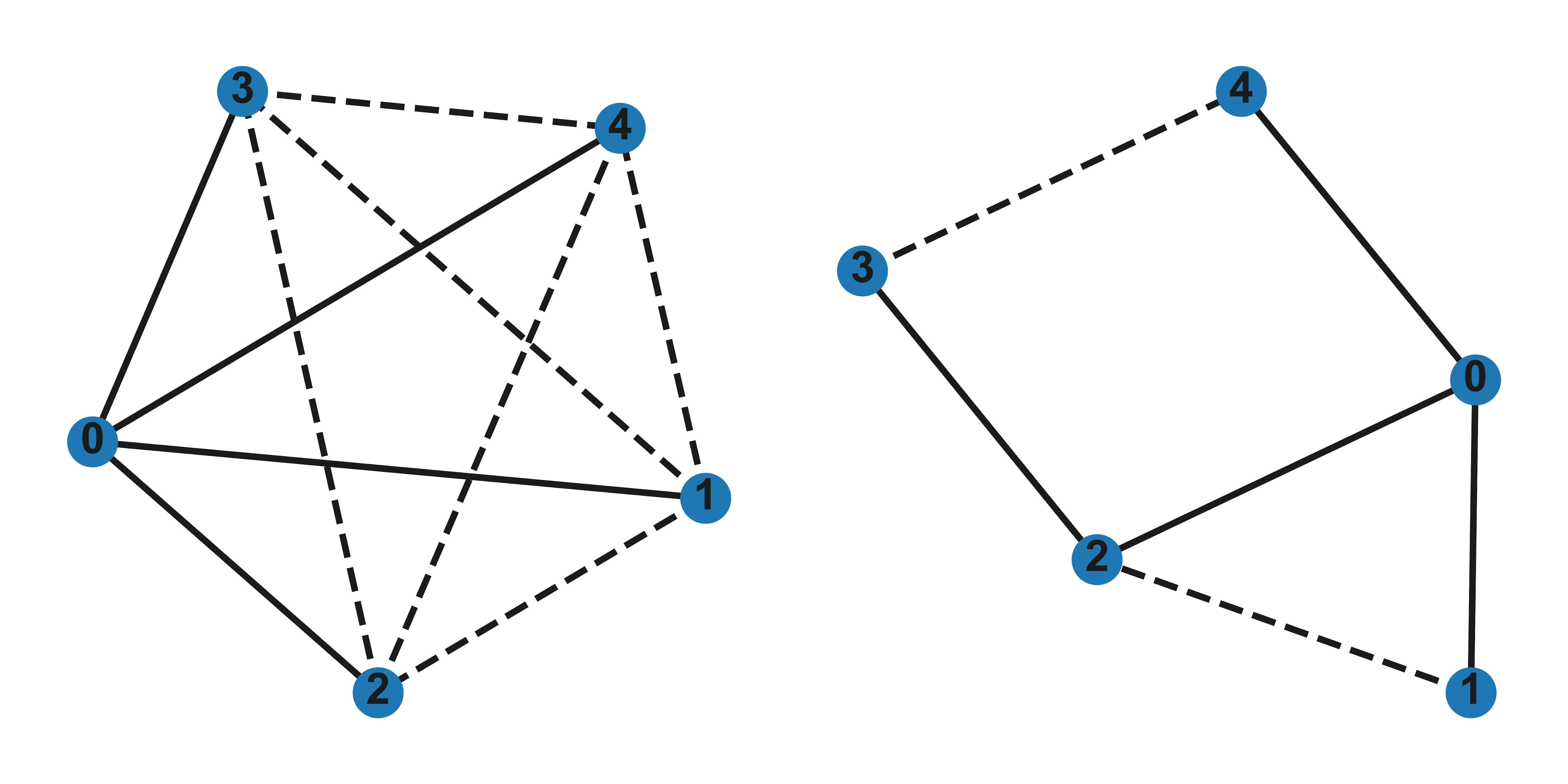}
\caption{Figure shows an array of five telescopes represented as vertices. The 
lines joining the vertices represent measured baselines. The graph on the
left is a complete graph with all ten baselines measured. The graph on the
right shows only six baselines measured. In both graphs, the solid lines 
form a spanning tree: a subgraph that contains all the vertices, but no   
closed  loops. The graph on the left gives us six advariants, while the graph on the right gives us
one advariant and one covariant.
}
\label{fig:SpanTree}
\end{figure}

       The left side of Figure \ref{fig:SpanTree} shows the case of five antennas as vertices of a connected graph. (A disconnected graph would require each connected piece to be treated separately). Since this has all 15 baselines as the edges, it  is a complete graph.  For the general complete graph with $N$ antennas, we can choose a base point and draw edges to each of the $(N-1)$ remaining antennas. That is shown in solid lines and forms a spanning tree -- a connected set of $(N-1)$  edges which visit all the $N$ vertices.    We thereby  get     $(N-1)(N-2)/2$ remaining edges, shown dotted,  each of which gives us one triangular circuit with respect to the base point. These are the advariants from which the complete set of invariants was constructed in \cite{Samuel}.

      To extend these ideas to the case of a connected incomplete graph, we proceed as follows. Given a graph ${\cal G}$, we pick a spanning tree -- a connected  subgraph ${\cal G}_0\subset {\cal G}$ which contains all the vertices but has no closed loops. 
      With $N$ antennas, this has $(N-1)$ edges. The spanning tree, by itself cannot give us any invariants,
      as it has no closed loops.
      For the same reason, there is a unique path in the spanning tree to travel from one vertex to any other. 
            
      The remaining  edges, not in the spanning tree, are  shown 
      as dotted lines in the figure. We now  put back these edges, each of which gives us a closed loop. This is obtained by travelling from the base point along the unique path  to one of the vertices of this edge, traversing the edge, and then returning to the base point, again by a unique path.  The final closed path from the base back to the base  may have an odd or an even number of edges. Each loop with an even number of edges give us a covariant and each loop with an odd number
      give us an advariant. 
      
      If there are no advariants, all the loops in ${\cal G}$ have an even length. In this special case,   one
      can consistently colour all the vertices of ${\cal G}$ alternately red and blue so that edges only connect
      vertices of different colours. This is the definition of a bipartite graph. In this case one only has 
      covariants. Invariants are produced by taking the trace and determinant of these covariants. If the the graph is not bipartite, there is at least one odd loop in ${\cal G}$. One can convert all the covariants into advariants by  concatenation -- first traversing the odd loop and then traversing any even loop.  Invariants are then constructed by the earlier method based on Lorentz transformations and four-vectors \cite{Samuel}. To summarise,  we only need to deal with two cases: one with only covariants (bipartite graph), and one with only advariants (non-bipartite graph).
      
      The right side of Figure \ref{fig:SpanTree} gives an example of an incomplete graph with five antennas and only six baselines measured.  A spanning tree --  not unique  -- for the graph is  shown using solid lines. 
      The remaining edges are shown dotted. From a given base point -- again not unique -- one can make one circuit for each of the dotted edges. The matrix products around  each of these loops are independent, since each contains one baseline not present in any other loop. These loops also form a kind of basis -- any other closed loop can be built up by traversing these basic loops successively. (More formally, these loops are {\it generators} of the fundamental group of the graph )

\section{Construction of  invariants}%

 Picking a common starting point for a complete set of circuits,  we now have  cancellation of the intermediate $\mathbf{G}$'s for  all the circuits. We are still left with  factors of  either a $\mathbf{G}$ and a
  $\mathbf{G}^\dagger$, or, a $\mathbf{G}$ and a $\mathbf{G}^{-1}$ at the two ends of a matrix product with alternating correlation matrices and their hatted forms. It now remains to construct scalar quantities from such a  product 
  which are independent of the terminal $\mathbf{G}$'s.  For advariants, the procedure  to construct invariants is 
   given in \cite{Samuel}. This relies on the connection between  $2\times 2$  matrices with determinant 1, and Lorentz transformations, long used in relativistic quantum theory. 
  This connection was first introduced in radio astronomy in the context of single dish polarimetry by \cite{Britton}.

    It is convenient, with  no loss of generality, to expand the matrix product around a closed loop in terms of the identity and the three  Pauli matrices, with four complex coefficients  $ a_0, a_1, a_2, a_3$. Premultiplying by $\mathbf{G}$ and postmultiplying by $\mathbf{G}^\dagger$ then produces a real Lorentz transformation among the four $a$'s and further scales them by a positive constant $\det\left(\mathbf{GG}^\dagger\right)$.   If there are only advariants, as in the complete graph case (\cite{Samuel}),  we construct a set of independent scalar products of these four vectors. Normalising (or taking ratios) to remove the overall scale is sufficient to give the desired number of invariants, The same  procedure works  for the incomplete graph case,  if we have only 
    advariants. As shown in \cite{Samuel}, with \(~N_A~\) advariants, we get  \(~8N_A-7~\) real  invariants. 
    If there is even a single advariant, one can traverse it and then any covariant, and the resulting product is an advariant.  By converting all covariants to advariants the earlier algorithm can be used. 
    
    The only case left is when there are only covariants, say $N_C$ of them. This is unlikely to occur in practice, but we present it below  for completeness.  All  covariants transform with a prefactor of $\mathbf{G}$ and a postfactor of $\mathbf{G}^{-1}$. A graph in which there are only covariants is necessarily a bipartite graph, as all closed loops have an even number of edges. In this case, $a_0$ remains invariant, since it is half the trace.  The determinant is also invariant, and it is $a_0^2 -a_1^2-a_2^2-a_3^2$ Therefore, the sum of the squares (not absolute squares!) of $a_1,a_2,a_3$ is separately an invariant, so we refer to it as a (complex) three-vector. For a single covariant, this approach  is equivalent to the determinant and trace, either way gives 4 real invariants. The traces by themselves contribute $2N_C$ real invariants, Setting these aside, with more than one  covariant, we can form  the complex  three dimensional scalar product   from the   three -vectors of two covariants, which is also  an invariant.   The first two covariants give us two complex vectors, which give us three complex  invariants, their squared lengths and inner product. The two  three-vectors are also enough to define a frame in this vector space, since we can take the third vector as their cross product. At this stage, with two covariants, we have six real invariants.  Each of the remaining $N_C-2 $ covariants now contributes six more real invariants, since one can construct scalar products with the three members of the frame. 
    So, for $N_C \geq 2$, we have  $2N_C +6+6(N_C-2) = 8N_C-6$ real  invariants in all, As in the case of advariants, $N_C=1$ is an exceptional case, with 4 rather than 2 invariants.

    \section{The role of invariants in imaging}%
\cite{Akiyama_2021} have used different strategies to produce the iconic image shown below.
The traditional route to determining and  refining
   the antenna based gains is self calibration (\cite{TMS}), in which, briefly, a model image and the gains are alternately updated. The gains are improved
   by fitting the model predictions  to the measurements, while the image based on the current gains is improved by  deconvolution.   This strategy bypasses invariants -- if the process converges, 
   then all invariants are automatically satisfied by the final image. 
   
   However, invariants are still in use and
   have independent value especially in cases when the data is not extensive, the results may depend significantly on the details of the initial guess 
   and the deconvolution algorithm. Accordingly, as recognised in the EHT work, there is a role for 
   `forward modeling' in which one computes the correlations from a parameterised  family of models. (For a recent simulation  of forward modeling  using machine learning  for parameter inference, see  \cite{nithyanew}). For example, in a  case like  M87, the shape of the ring,
   the depth of the dark region inside it could be parameters.  One can then compare the predicted invariants with those computed from the measured
   correlations. We have explored the use of invariants in this approach. Two questions need to be answered. a)  given the presence of inverse matrices
   in the covariants and advariants, could  near  singular behaviour be an obstacle? b) Given that the invariants do not have an intuitive geometric meaning
   even in the single polarisation case, what features of the model can be reliably determined from them in a forward modeling framework?

\section{Simulations with noise }%

  A set of simulations was carried out on a toy version of the EHT $u-v$ coverage. 
  The polarised emission from the source was modelled as a ring with the thickness arising out of a gaussian intensity profile. The intensity also has a sinusoidal variation along the azimth of the ring. This intensity distribution is shown in red in Figure \ref{Rings}. The yellow lines in the figure indicate the direction of polarisation. The free parameters related to the polarisation are the degree of linear polarisation (constant everywhere over the ring) and the azimuthal variation of the polarisation direction. The degree of circular polarisation is taken to be zero.

  A snapshot based on $u-v$ values for 
   five locations - Hawaii, Chile, Mexico,  two in the US mainland, and Spain - was used. The signal to noise  was taken to be 100 for the strongest correlation and the same noise amplitude used for the
   weaker ones. Invariants were computed using the prescription in \cite{Samuel}. The publicly available python package, NetworkX (\cite{hagberg2008exploring}), was used to consturct the graphs, find spanning trees and compute invariants. A violin plot of the invariants is shown below, based on 5000 realisations of noise on the correlations.

   \begin{figure}
    \centering
    
     \includegraphics[width = \columnwidth]{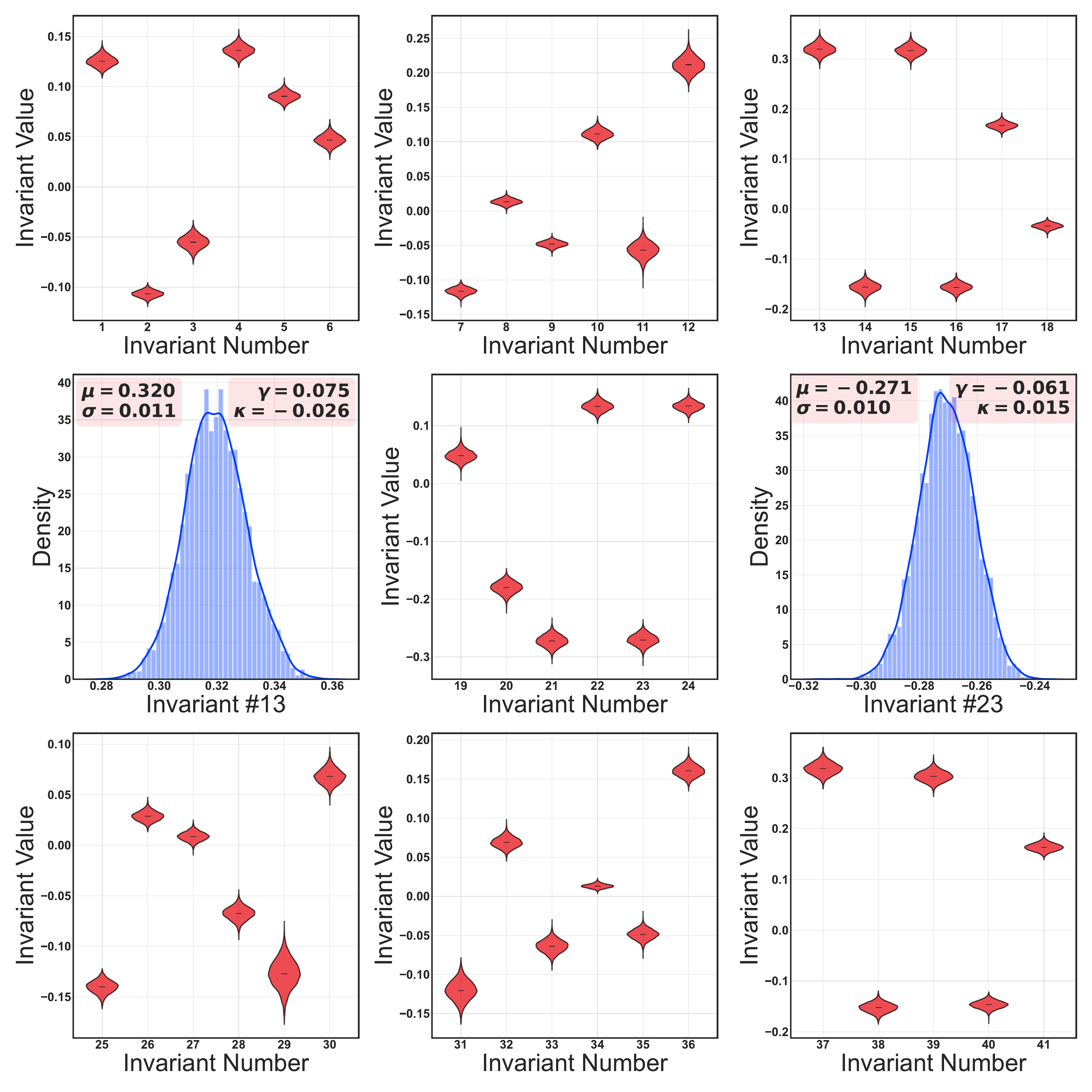}
    
    \caption{Violin plots showing the distribution of the 41 invariants for the complete graph case. The two bar plots show the distribution of two of the invariants in detail. The mean, standard deviation, skewness and kurtosis of these distributions are listed in the top right corners of the respective plots. The non-zero values of the skewness and the kurtosis indicate the non-gaussianities of the distribution.  }
    \label{fig:ViolinComplete}
    \end{figure}

   The violin plots in Figure \ref{fig:ViolinComplete} gives a feel for the noise level.  While the signal to noise on some of the invariants is poor, it is clear that there is no major issue with the inverse matrices which occur in the construction of the invariants. This can be rationalised  from the fact that complex $2\times2$  matrices form an 8 dimensional space, while the singular ones form a six  dimensional space (from the vanishing of a complex determinant).

\begin{figure}
    \centering
    
     \includegraphics[width = \columnwidth]{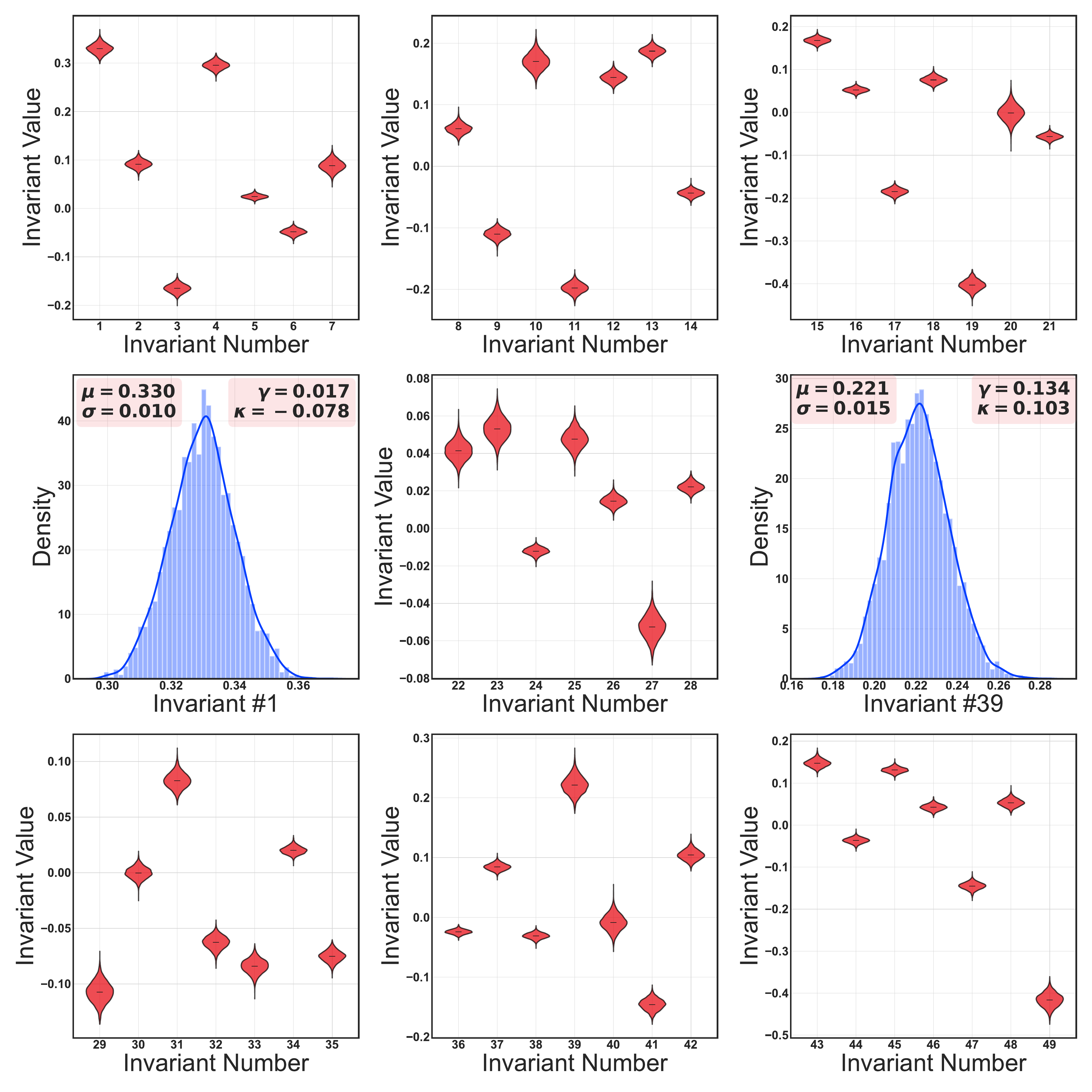}
    
    \caption{Violin plots showing the distribution of the 49 invariants for the incomplete graph case. As before, the two bar plots show the distribution of two of the invariants in detail, along with the mean, standard deviation, skewness and kurtosis.  }
        \label{fig:ViolinInComplete}
    \end{figure}

   A similar exercise with an incomplete graph was carried out. This graph had 6 antennas in total, placed at randomly chosen points on the $u-v$ plane. Out of the 15 possible correlations, 12 were considered available. Noise properties were assumed to be the same as in the complete graph case. The invariants and their distributions are shown in Figure \ref{fig:ViolinInComplete}.

   This already shows in a quasi realistic case, that the invariants constructed from advariants following the recipe in \cite{Samuel} do not have a major problem in the presence of noise.
  However, in forward modeling, an important role is played by the likelihood as a function of the parameters (see appendix to \cite{Nithya}, which argues for  independence of the 
  precise choice of invariants). For each set of parameter 
  values, the likelihood  has to be computed anew. Our simulations clearly show that   some of the invariants have non gaussian distributions, and are clearly correlated, as shown by the computed covariance matrix. Computing a more realistic likelihood function is therefore  a challenge  which we are exploring. 

  \section{Acknowledgements}
   We acknowledge support of the Department of Atomic Energy, Government of India, under project no. RTI4001. J. S. and R. N. acknowledge support by a grant from the Simons Foundation (677895, R.G.). We thank Nithyanandan Thyagarajan for sharing his work with us before publication, and for a lively discussion.

\bibliographystyle{apsrev4-2}

\bibliography{main}

\end{document}